\documentclass[12pt,preprint]{aastex}

\input epsf.sty

\slugcomment{To appear in the Astrophysical Journal (October 20, 2004)}

\def\lsim{\mathrel{\rlap{\lower 4pt \hbox{\hskip 1pt $\sim$}}\raise 1pt \hbox
        {$<$}}}
\def\gsim{\mathrel{\rlap{\lower 4pt \hbox{\hskip 1pt $\sim$}}\raise 1pt \hbox
        {$>$}}}

\newcommand{\etal}{et~al.\ }
\newcommand{\eg}{e.g.,\ }
\newcommand{\ie}{i.e.,\ }
\newcommand{\Msun}{M_{\odot}}

\newcommand{\kms}{km~s$^{-1}$}

\newcommand{\OI}{O~{\sc i}}

\newcommand{\SiI}{Si~{\sc i}}

\newcommand{\SI}{S~{\sc i}}
\newcommand{\CaII}{Ca~{\sc ii}}

\newcommand{\FeII}{Fe~{\sc ii}}
\newcommand{\FeIII}{Fe~{\sc iii}}

\newcommand{\Cofs}{$^{56}$Co}
\newcommand{\Nifs}{$^{56}$Ni}
\newcommand{\Mej}{$M_{\rm ej}$}
\newcommand{\KE}{$E_{\rm kin}$}
\newcommand{\vph}{v_{ph}}

\begin{document}

\title{Properties of Two Hypernovae Entering the Nebular Phase:
    SN~1997ef and SN~1997dq}
\author{P. A. Mazzali\altaffilmark{1,2,3}, J. Deng\altaffilmark{1,2},
K. Maeda\altaffilmark{2}, K. Nomoto\altaffilmark{1,2},
A. V. Filippenko\altaffilmark{4}, T. Matheson\altaffilmark{5}}

\altaffiltext{1}{Research Center for the Early Universe, University of Tokyo,
    	Bunkyo-ku, Tokyo, Japan.}
\altaffiltext{2}{Department of Astronomy, University of Tokyo, Bunkyo-ku, Tokyo, 
Japan.}
\altaffiltext{3}{Osservatorio Astronomico, Via Tiepolo, 11, 34131 Trieste,
    	Italy.}
\altaffiltext{4}{Department of Astronomy, University of California, Berkeley, 
	CA 94720-3411.} 
\altaffiltext{5}{Harvard-Smithsonian Center for Astrophysics, 60 Garden St., 
	Cambridge, MA 02138.}
	
\begin{abstract}

The two peculiar Type Ic supernovae (SNe) 1997ef and 1997dq are shown to have
very similar photometric and spectral evolution in the epochs when both SNe are
observed (\ie beyond $\sim 80$ days after the explosion). The early light
curves and spectra of SN~1997ef suggested that this was a ``hypernova,'' or
``SN 1998bw-like Type Ic supernova.''  The fact that the two SNe are very
similar allows us to extend the time coverage of this type of event, since
SN~1997dq, unlike SN~1997ef, was observed well into the nebular phase. In
contrast to SN~1998bw, the spectra of these two SNe did not become fully
nebular until almost one year after the explosion. During a long transition
phase, lasting at least 6 months, the SNe developed nebular emission in lines
of [\OI] and [\CaII], but at the same time they retained an underlying,
photospheric-type spectrum, originating at very low velocities.  Spectral
synthesis techniques are used to model the spectrum of SN 1997dq, suggesting
that it produced $\sim 0.16~ \Msun$ of \Nifs, and that a significant fraction of
this is located in a dense, low-velocity inner region.

\end{abstract}


\section{Introduction}

The discovery of SN~1998bw, probably associated with GRB 980425 (\eg Galama
\etal 1998), provided the first direct evidence that supernovae (SNe) and
gamma-ray bursts (GRBs) may be related. Analysis of the SN~1998bw data showed
that this was a peculiar Type Ic supernova (SN~Ic; see Filippenko 1997 for a
summary of SN classifications), with a very large explosion energy, some 30--50
times the energy of an ordinary core-collapse SN ($10^{51}$ erg). For this
reason, SN~1998bw was called a ``hypernova'' (Iwamoto \etal 1998; Nakamura
\etal 2001).

   While an effort was made to establish whether other SNe were linked to GRBs
(Wang \& Wheeler 1998), additional SNe were studied which show the typical
characteristics of ``hypernovae,'' namely a high kinetic energy per unit mass,
exemplified by broad spectral features; see Mazzali \etal (2003a), and
references therein. Since not all of them are extremely luminous, but are
spectrally similar to SN~1998bw, they have also been dubbed ``SN 1998bw-like
SNe~Ic'' (Foley et al. 2003).

Analysis of one of these objects in particular, SN~1997ef, showed that its
kinetic energy is $\sim 2 \times 10^{52}$ erg, less than the energy of
SN~1998bw but still very large (Iwamoto \etal 2000; Mazzali \etal
2000). SN~1997ef was estimated to have synthesized $\sim 0.13~ \Msun$ of \Nifs,
or about twice the amount produced by normal core-collapse SNe, although again
less than that of SN~1998bw. A clear connection with a GRB could not be
established for SN~1997ef, yet a possible candidate was found (GRB 971115; Wang
\& Wheeler 1998). The common features of SNe 1997ef and 1998bw are the broad
absorption troughs in the spectra near maximum brightness, indicating the
presence of high-velocity ejecta, and the luminous and extended light curve. 
Other examples of SN~1998bw-like SNe~Ic are SN~2003dh/GRB030329 (Stanek \etal 2003; 
Hjorth \etal 2003; Matheson \etal 2003; Mazzali \etal 2003b; Kawabata \etal 
2003), which was very similar to SN~1998bw; SN~2002ap (Mazzali \etal 2002; 
Leonard \etal 2002; Foley \etal 2003), which was less energetic and is 
thought to have synthesized only $\sim 0.08~ \Msun$ of \Nifs; and recently 
SN~2003lw/GRB031203 (Tagliaferri \etal 2004; Malesani \etal 2004).

The study of SN~1998bw extended to late times. The spectrum turned nebular
after $\sim 100$ days, while the light curve showed a phase of exponential
decline. This can be interpreted as being due to the presence of an inner
high-density region (Nakamura \etal 2001; Nomoto \etal 2001; Maeda \etal
2003). The fully nebular spectrum of SN~1998bw showed broad lines, with the
peculiarity that the oxygen lines were narrower than the iron lines. This is
difficult to interpret in the context of a 1-dimensional, spherically symmetric
model, but it can be explained if the explosion was asymmetric (Mazzali \etal
2001; Maeda \etal 2002). On the contrary, SN~1997ef retained a
photospheric-type spectrum, with absorption lines of progressively lower
velocity, up to the last previously studied spectrum, dating $\sim 2$ months
after the explosion (Mazzali \etal 2000), when some evidence of nebular
emission is also present. The long duration of the photospheric phase can also
be interpreted as the consequence of the presence of an inner, high-density
region, but in this case the density contrast between such a region and the
outer ejecta may be larger (Maeda \etal 2003).

In this paper, we analyze spectra of SN~1997ef extending beyond the previous
study (to epochs of $\sim 3$--4 months), and show that at comparable epochs,
while SN~1998bw had a fully nebular spectrum, SN~1997ef still retains a
photosphere.  We study these spectra with a double-barreled approach, dividing
them into a photospheric and a nebular part, and obtain an estimate of the mass
of \Nifs\ powering the spectrum at these intermediate epochs.  Unfortunately,
no good spectra of SN~1997ef exist at later times. However, another peculiar
SN~Ic, SN~1997dq, was shown by Matheson \etal (2001) to be very similar to
SN~1997ef. Combining the light curves of these two SNe we generate a
``SN~1997ef-like'' extended light curve.  This is particularly useful because
SN~1997dq does have a good-quality spectrum obtained in the fully nebular epoch
($\sim 300$ days). This spectrum can thus be used to determine directly the
mass of \Nifs\ produced by SN~1997dq and, by analogy, by SN~1997ef.  This can
be compared to the result obtained from the light curve near peak.

\section{The Intermediate-Epoch Spectra of SN~1997ef}

SN~1997ef, a peculiar SN~Ic, was classified as a hypernova both by Branch
(2001), who found a kinetic energy of $\sim 3 \times 10^{52}$ erg from spectral
fits, and by Iwamoto \etal (2000), who derived from a simultaneous study of the
light curve and the spectra the values \KE $= 8 \times 10^{51}$ erg, \Mej $=
8~\Msun$, and $M$(\Nifs)$ = 0.15~\Msun$. Iwamoto \etal suggested that SN 1997ef
was the result of the collapse of the stripped CO core of a star having a
zero-age main sequence (ZAMS) mass of $\sim 30$--35$\Msun$.  Following the
probable association between SN~1998bw and GRB 980425, Wang \& Wheeler (1998)
looked for other possible associations between SNe and GRBs, and suggested that
GRB 971115 may be compatible with SN~1997ef in position and time of occurrence,
although the correlation is much weaker than for SN~1998bw and GRB 980425 and
thus the link between the SN and the GRB cannot be proved.

Mazzali \etal (2000) analyzed a series of photospheric-epoch spectra of
SN~1997ef.  Using spectroscopic dating on the 1997 Nov. 29 spectrum (UT dates
are used throughout this paper), they estimated that the explosion occurred on
Nov. 20 $\pm 1$ day.  They derived the density and abundance distributions in
the ejecta, and found that typical SN~Ic abundances describe SN~1997ef very
well, but their results differed from the explosion model used to describe the
light curve in two ways: (1) the outer density structure turned out to be
rather flat (a power law with index $n = -4$), so that the very broad
absorption features, extending to $v \approx 25000$ \kms, could be reproduced,
and (2) matter was found to be present at low velocities ($\sim 2000$ \kms),
which is well below the low-velocity cutoff of the spherically symmetric
explosion model (5000 \kms).  Their most important conclusion is that a
hypernova model with \KE$ = 2 \times 10^{52}$ erg, \Mej$ = 10~\Msun$, and
$M$(\Nifs)$ = 0.13~\Msun$ (model CO100) gives a good representation of the light
curve and the spectra.

A synthetic light curve obtained with this modified density structure fits the
observations near the peak quite nicely, and Mazzali \etal (2000) suggested
that if an additional $0.05~ \Msun$ of \Nifs\ is contained at low velocities and
high densities, then the observed slowing down of the light curve between days
60 and 160 might be explained by the ensuing additional deposition of
$\gamma$-rays. They also suggested that the distribution of \Nifs\ may be
aspherical.

A more conclusive determination of the \Nifs\ mass could be obtained from the
nebular spectra. Mazzali \etal (2000) only extended their study to the first
$\sim 2$ months. The last spectrum they analyzed, dated 1998 Jan. 26, has an
epoch of 67 days after the explosion.  Superposed on the photospheric spectrum,
this spectrum showed net emission in [\OI] $\lambda$6300, [\CaII]
$\lambda$7300, and the \CaII\ near-infrared triplet, which is typical of
SNe~Ic.

Matheson \etal (2001) presented spectra of SN~1997ef obtained at later epochs.
Only two of these spectra can be used for modelling: those observed on 1998
March 7 and March 26. The 1998 Sep. 21 spectrum is unfortunately too noisy,
even after smoothing or rebinning, and it has almost no signal in the
blue. According to the spectral dating of Mazzali \etal (2000), the explosion
took place on 1997 Nov. 20; hence, the two useable spectra have epochs of
107 and 126 days, respectively. At such epochs, the spectra of SN~1998bw were
essentially fully nebular. In contrast, for SN~1997ef this is not the
case: despite the greatly increased strength of the emission lines, a
comparison with the 1998 Jan. 26 spectrum shows that both March spectra are
still dominated by P-Cygni profiles superposed on a cool continuum, which is
almost flat in the optical.  Thus, the March spectra can be divided into two
components, one photospheric and one nebular.

Because it is interesting to determine the properties of both components, we
modelled the March spectra using a double-barrelled approach: we reproduce the
photospheric component using our Monte Carlo code (Mazzali 2000), which gives
us information on the luminosity, the photospheric velocity, and the
abundances, but is not able to reproduce net nebular line emission. This is
then reproduced with our non-LTE (NLTE) nebular spectral code (\eg Mazzali
\etal 2001), from which the \Nifs\ mass contributing to powering the nebular
spectrum can be determined.  The sum of the two component spectra should
reproduce the observations.

We have therefore adopted the following modelling procedure.  First, the
spectra were calibrated according to the $V$ magnitude obtained by
interpolating the light curve.  Second, the spectra were modelled with our
photospheric-epoch Monte Carlo code.  Finally, the remaining net emission was
modelled using our NLTE nebular code.  The main results are outlined in the
next two sections.

\subsection{1998 March 7, Day 107}

The photospheric component requires a cool spectrum. The Monte Carlo model
requires $L = 4.8 \times 10^{41}$ erg s$^{-1}$ and $v_{ph} = 1650$ \kms.  This,
combined with the epoch, gives $T = 4200$~K, and the synthetic spectrum has
$V = 18.70$ mag, $B = 19.97$ mag, and bolometric correction (BC) = $-0.57$ mag.
Oxygen, silicon, and sulfur dominate the ejecta. The iron abundance is small.

The synthetic spectrum reproduces reasonably well the blue part of the observed
spectrum ($\lambda < 6000$~\AA), and the P-Cygni lines of \OI\ $\lambda$7770,
the \CaII\ near-IR triplet, and \SI\ $\lambda$9220. In the synthetic spectrum,
the \FeII\ feature near 5000~\AA\ is too deep, and the \CaII\ near-IR triplet
absorption does not extend far enough in the blue, probably because of our
model grid.  The emission features which are not reproduced are as mentioned
above for the 1998 Jan. 26 spectrum (\ie [\OI] $\lambda$6300, [\CaII]
$\lambda$7300, and the \CaII\ near-IR triplet), plus a weak line near 5550~\AA\
(probably [\OI] $\lambda$5577).

%*** Paolo, I explicitly mentioned the emission lines that don't fit well,
% in the line above. Are these the right ones?
% Also: clarify what you mean when you say that the ``model grid'' may be
% responsible for the poor fit of the Ca II triplet absorption.

We modelled the residual emission spectrum trying to fit the strongest
uncontaminated lines, [\OI] $\lambda$6300 and [\CaII] $\lambda$7300. The lines
are rather narrow, requiring an outer velocity of the nebula of 5000 \kms. Note
that at a comparable epoch SN~1998bw had line half-widths of $\sim 12000$ \kms\
in both [\OI] and [\FeII]. It is not possible to isolate [\FeII] emission in
SN~1997ef, so an estimate of the \Nifs\ mass must rely on fitting the [\OI]
line and not exceeding the total observed flux in the [\FeII] region.

We found that a nebula containing $0.05~\Msun$ of \Nifs, $0.5~\Msun$ of O,
$0.4~\Msun$ of Si, and smaller masses of other elements gives a good fit to the
observed residual emission spectrum. The line at 6500~\AA\ is reproduced as
[\SiI] $\lambda$6527.

Significant clumping (volume filling factor $v_f = 0.1$) is adopted in analogy
with SN~1998bw (Mazzali \etal 2001). This is necessary to keep the ionization
sufficiently low that [\FeIII] lines do not form.

The luminosity of the nebular spectrum is $9 \times 10^{40}$ erg s$^{-1}$.
This luminosity corresponds to a $\gamma$-ray deposition fraction
$D_{\gamma} = 0.38$ for a \Nifs\ mass of $0.05~\Msun$.

Figure 1 shows the observed spectrum, the Monte Carlo synthetic spectrum, the
nebular spectrum, and the sum of the latter two. The combined synthetic
spectrum reproduces the observations well, including an improved fit in the
5000~\AA\ region. Regions of low line opacity ($\sim 6800$~\AA, $> 9600$~\AA)
are not well reproduced because our Monte Carlo model assumes a sharp
photosphere as the lower boundary.  The total luminosity of the spectrum is
then $L = 5.7 \times 10^{41}$ erg s$^{-1}$, or $M({\rm Bol}) = -15.64$ mag.

The luminosity of the photospheric component alone corresponds to the complete
deposition of $\gamma$-rays and positron from $0.10~ \Msun$\ of \Nifs,
indicating that this is the minimum value of the \Nifs\ mass located below
$\vph$. The real value is probably larger, as many of the $\gamma$-rays
probably escape. Thus, the total mass of \Nifs\ is at least $0.15~ \Msun.$

\subsection{1998 March 26, Day 126}

The differences between this spectrum and the previous one are small. A
photosphere is still present. The depth of the absorption in the \FeII\
region is reduced, and the relative importance of net emission is greater,
including now clearly [\OI] $\lambda$5577.

The Monte Carlo model requires $L = 4.1 \times 10^{41}$ erg s$^{-1}$ and
$v_{ph} = 1350$ \kms.  We obtain $T = 4100$~K, and the spectrum has $V=18.93$
mag, $B = 20.14$ mag, and $BC = -0.63$ mag. The elements O, Si, and S dominate
the ejecta. The Fe abundance is smaller than on March 7.  The fit is similar to
that of March 7.

The nebula has $v = 5000$ \kms, and it contains $0.05~\Msun$ of \Nifs,
$0.7~\Msun$ of O, $0.4~\Msun$ of Si, and smaller masses of other elements.
Clumping is again assumed. The nebular spectrum has $L = 7 \times 10^{40}$ erg
s$^{-1}$, \ie $D_{\gamma} = 0.35$. The O mass changes significantly in the two
models, while other masses do not. Unfortunately, determinations of the O mass
are intrinsically very uncertain (\eg Mazzali \etal 2001), and the values
derived here should be regarded as only approximate.  The combined synthetic
spectrum is a reasonably good fit to the observed one (Figure 2).

The total luminosity of the spectrum is $L = 4.8 \times 10^{41}$ erg s$^{-1}$,
\ie $M(\rm Bol) = -15.50$ mag. The luminosity of the photospheric component
corresponds to the complete deposition of $\sim 0.1~ \Msun$\ of \Nifs,
confirming that the total mass of \Nifs\ should be at least $0.15~ \Msun$.

\section{Light Curve}

In Figure 3 we show the new extended light curve of SN~1997ef. The
``bolometric'' light curve was derived from the model fits, not from
integrating the spectra.  The BC changes from negative to positive on about day
60, and the bolometric light curve is brighter and somewhat flatter than the
$V$ light curve at advanced epochs. The bolometric light curve between 1998
Mar. 7 and Mar. 26 (days 107--126) is only slightly steeper than the \Cofs\
decay slope. The luminosity for Jan. 26 (day 67) is the result of modelling the
photospheric spectrum only, and it does not include net emission which is
present. Its value is therefore somewhat underestimated.

The late-time bolometric light curve of SN~1997ef appears to follow the decay
rate of \Cofs. If we assume that the deposition of the $\gamma$-rays and
positrons is complete ($D_{\gamma} = 1$) in this late phase, a \Nifs\ mass of
$0.10~ \Msun$ is derived. However, this must be an underestimate, because $0.13~
\Msun$ of \Nifs\ was required to explain the peak of the light curve, and the
subsequent decline was faster than the \Cofs\ slope, so that $D_{\gamma}$,
integrated over the entire ejecta, must be less than 1 in the late phase.  In
the combined photospheric and nebular models for SN~1997ef, we find $D_{\gamma}
\approx 0.35$ in the nebula. If we take this value of $D_{\gamma}$ to be
representative of the entire SN, we require $M$(\Nifs)$ = 0.29~\Msun$ to fit the
luminosity, which is certainly an overestimate, because there is still a
photosphere in the March 1998 spectra.  This suggests that $M$(\Nifs) may be
about $0.2~\Msun$, part of which is buried deep in the ejecta so that
$D_{\gamma} \approx 1$, and the rest is in the nebula where $D_{\gamma}$ would
change with radius or position in a more continuous manner.

The presence of \Nifs\ deep in the ejecta, which contributes only marginally to
the peak of the light curve, may be in agreement with the observed presence of
material at velocities below the cutoff required for the ejection of $0.13~
\Msun$ of \Nifs. However, this puzzling finding also suggests that the
explosion may not be described accurately if simple spherical symmetry is
adopted.

The combination of the photospheric and nebular components in the
advanced-phase spectra of SN~1997ef seems to require a value of the \Nifs\
mass of at least $0.15~ \Msun$, slightly larger than the estimate from the
peak of the light curve. This value is subject to uncertainties because of
the incomplete $\gamma$-ray deposition. A more accurate determination of the
\Nifs\ mass may be possible by analyzing data from later epochs, when the SN
should more closely approximate a nebula. Unfortunately, the only spectrum of
SN~1997ef which is sufficiently late is also very noisy, and only the [\OI] and
\CaII\ lines can be detected. However, Matheson \etal (2001) suggested that
another SN~Ic, SN~1997dq, is a close analogue of SN~1997ef, and there exist
nebular spectra of this SN.

\section{SN~1997dq: A SN~1997ef-Like Supernova}

Matheson \etal (2001) showed a set of spectra of the peculiar SN~Ic 1997dq and
suggested that it is similar to SN~1997ef (cf.  their Figures 17 and 18).  In
Figure 4 we compare the normalized light curves of these two SNe. The light curve
of SN~1997dq (Nakano \& Aoki 1997; Moretti \& Tomaselli 1997, 1998a,b,c) is
shifted by 1.8 mag and the epoch is defined so that the two light curves
overlap. These light curves seem to be very similar, suggesting that the
properties of SN~1997dq were either the same as those of SN~1997ef or fell
along the track of parameters (\Mej, \KE) which produce equivalent light
curves.  However, having assigned an epoch to the spectra of SN~1997dq on the
basis of the light-curve comparison, its spectra can now be compared with those
of SN~1997ef taken at a nominally similar epoch. This comparison (Figure 5)
shows that the two SNe also have very similar spectra. Because the spectra
(line width and velocity) also depend on \Mej\ and \KE, but do so in a different
way than does the light-curve shape, the fact that both the light curves and
the spectra of the two SNe are similar shows rather conclusively that they had
comparable parameters. The only slight uncertainty may be in the mass of \Nifs,
because this affects the overall brightness of the light curve, but not so much
its shape. An independent determination of the mass of \Nifs\ can, however, be
obtained from the nebular spectrum.

As we mentioned above, a nebular spectrum is available of SN~1997dq. This can
be used to determine the mass of \Nifs\ ejected by this SN and, by analogy, by
SN~1997ef. The spectrum was obtained on 1998 June 18 (Matheson \etal 2001).
Given the dating of the SN~1997dq light curve based on the comparison with the
light curve of SN~1997ef, the epoch of this spectrum is 262 days after
explosion.  A small error in the determination of the epoch does not affect the
results of the modelling.

We modelled the 1998 June 18 spectrum with our NLTE nebular code (\eg Mazzali
\etal 2001).  Given the relative distance modulus of SN~1997dq with respect to
SN~1997ef ($-1.8$ mag), we adopted for SN~1997dq a distance modulus $\mu =
31.83$ mag ($d \approx 23$ Mpc).  We assumed that no reddening is present,
as for SN~1997ef.

The best fit (Figure 6) required $M$(\Nifs) $= 0.16~ \Msun$, and a total mass
of $2.3~ \Msun$, of which only about $0.8~ \Msun$ consists of O, the rest being
mostly Si, C, and S. The masses of Ca and Mg are rather small by comparison,
even though they do give rise to rather strong emission features. The nebula
has an outer velocity of only $\sim 5000$ \kms, which is similar to the
velocity of the nebular component of the spectra of SN~1997ef. The fact that
the velocities of the regions that form the nebular spectrum are low confirms
the finding of Mazzali \etal (2000) that significant amounts of mass are
present at velocities lower than the cutoff imposed by 1-dimensional explosion
models. Indeed, Mazzali \etal (2000) estimated that the mass between 3000 and
5000 \kms\ is $1.65~ \Msun$, in good agreement with our result here, which
refers to all regions with $v < 5000$ \kms.  The density of the outer regions,
which contributed to the early spectra, is too low at these advanced epochs to
contribute significantly to the nebular spectrum, and the deposition of
$\gamma$-rays and positrons is also probably quite small, because \Nifs\ is not
expected to be mostly located in those outer regions. The estimate of the
\Nifs\ mass is also in good agreement with the results obtained for SN~1997ef.
Although the spectrum is not of exceptionally good signal-to-noise ratio, the
line profiles seem to show no signs of an asymmetric explosion, such as the
double components, narrow and broad, observed in the nebular spectra of
SN~1998bw (Mazzali \etal 2001).

The only available photometry of SN~1997dq are the unfiltered CCD points
reported in several IAUCs. Since we do not know the exact response of the CCDs
that were used to observe SN~1997dq,  the $V$-band fluxes derived from the
unfiltered CCD fluxes to compare SN~1997dq to SN~1997ef have an intrinsic
uncertainty.  CCD's typically have a response similar to that of the $R$-band
(see, e.g., Sec. 2 and Fig. 27 in Matheson et al. 2001).  The spectra of
SN~1997dq show that its $R$ and $V$ magnitudes have a difference of $\sim 0.5$
mag.  We take this as an estimate of the uncertainty of approximating the $V$
flux to the CCD flux. To account for this uncertainty, we also modelled the
nebular spectrum using distance moduli of 32.33 and 31.33 mag. We can still
obtain good fits of the nebular spectrum using the same outer velocity for the
nebula, but the \Nifs\ mass and the total ejecta mass increase to $0.195~\Msun$
and $2.7~\Msun$, respectively, if the larger distance is used, and decrease to
$0.135~\Msun$ and $2.0~\Msun$, respectively, if the shorter distance is used. 
This range of values (in particular the mass of \Nifs) is consistent with the
SN having produced more \Nifs\ than the average SN~Ic.

The main discrepancies between the synthetic and observed spectra are the
emission near 7750\AA, which is absent in the model, and the
[\CaII]~7291,7324\,\AA, which is too broad in the model. As for the first
feature, it is tempting to attribute it to \OI~7775\,\AA. However, this line is
very weak in our models, as it comes from rather highly excited upper levels.
Signs of a narrow \CaII~IR triplet are also seen, and this could indicate an
inhomogeneous distribution of Ca. However, the 7291,7324\,\AA\ line is also
affected by [\FeII] emission ($\lambda \lambda 7155$,7388), which contributes
to making the synthetic feature too broad. A three-dimensional study, or at
least a one-dimensional stratified one, may be required to understand this
behaviour. Both problems, however, may simply be due to our still rather
limited knowledge of Fe-group lines collision strengths.

In conclusion, our simulations of the nebular spectrum of SN~1997dq confirm
that this SN had very similar properties to SN~1997ef, and in particular that
it ejected $\sim 0.16 \pm 0.03~ \Msun$ of \Nifs. The final points in the light
curve of SN~1997ef are indeed brighter than the synthetic light curves, which
were computed for $0.13~ \Msun$ (Mazzali \etal 2000).

\section{Discussion}

Using the advanced-phase spectra of SN~1997ef and one nebular spectrum of
SN~1997dq we have shown that the two SNe are very similar, so that SN~1997dq
can also be dubbed a hypernova. We also found that the mass of \Nifs\ ejected
by both SNe is $\sim 0.16~ \Msun$, somewhat larger than the $0.13~ \Msun$
necessary to fit the early light curve.

The fact that SN~1997ef retained a photospheric component to its spectrum up to
at least day 126 suggests that even though the outer ejecta were expanding at
high velocities, a significant fraction of the mass was ejected at very low 
velocities. This contradicts the results of 1-dimensional explosion
calculations (Mazzali \etal 2000), which produce ejecta with a well-defined 
lower velocity boundary, but is consistent with the finding by Mazzali
\etal (2000) from modelling the early-time spectra that only $\sim 0.04~ \Msun$
of \Nifs\ is located at $v > 4000$ \kms.

Maeda \etal (2003) computed a light curve with our Monte Carlo model based on
the above prescriptions. They find that the increased \Nifs\ mass helps to 
reproduce the late-time light curve, but it does not affect the early-time
light curve if it is contained in an inner region of low velocity and high
density which is added to the original density structure. This inner region is
also necessary to have the prolonged photospheric-type spectra.

A possibility is that SN~1997ef, like other SN~1998bw-like SNe~Ic, was a
highly aspherical explosion. Two-dimensional calculations of jetted explosions
show that the asphericity significantly alters the density and Ni
distributions, especially in the central region (Maeda \etal 2002; Maeda \&
Nomoto 2003).  Signatures of asphericity, however, are not easy to detect,
partly due to the paucity of detailed and direct theoretical predictions by
multi-dimensional radiation calculations. If SN~1997ef was a highly aspherical
explosion, it is likely that the SN was not viewed from very close to the jet
axis, since the case for association with a GRB is not strong.

It is interesting that at least 2 objects out of the 28 SNe~Ib/Ic discussed by
Matheson \etal (2001) are SN~1998bw-like SNe~Ic. SNe~Ib/c are thought to
originate from the stripped CO cores of massive stars (\eg SN~1994I had a core
mass $M_{CO} \approx 2~ \Msun$ and a main-sequence mass of $\sim 15~ \Msun$;
Iwamoto \etal 1994). For SN~1997ef, the ZAMS mass was estimated to be $\sim
30$--35$\Msun$. The least massive SN~1998bw-like SN~Ic known to date is
SN~2002ap, for which the ZAMS mass was estimated at $\sim 20$--25~$\Msun$.  For
a normal initial mass function, $\sim 20$\% of all stars with $M_{ms} > 8~
\Msun$ have $M > 25~ \Msun$.  Podsiadlowski \etal (2004) have recently
suggested that not all massive, stripped-envelope stars become SN~1998bw-like
SNe (although all long-duration GRBs may be SN~1998bw-like SNe).

There may be other objects in the Matheson \etal (2001) sample which are SN
1998bw-like SNe~Ic. In particular, judging from the nebular velocities at
epochs of $\gsim 100$ days, Table 5 of Matheson \etal (2001) suggests that
SNe 1995F and 1990B may have had energies larger than the lower-energy SN~Ic
1994I. It would be interesting to study the properties of a sample of SNe~Ib/Ic
to determine their kinetic energies and the amount of \Nifs\ they synthesized,
in order to verify whether they might populate the plots relating these
quantities to the ZAMS mass (Nomoto \etal 2003).

\bigskip

A.V.F.'s research is supported by National Science Foundation grant
AST-0307894.  K.N.'s work has been supported in part by the Grant-in-Aid for
Scientific Research (14047206, 14540223, 15204010) of the Ministry of
Education, Science, Culture, Sports, and Technology in Japan.

%%%%%%%%%%%%%%%%%%%%%%%%%%%%% figures %%%%%%%%%%%%%%%%%%%%%%%%%%%%%%%%%%%
\newpage

%%%%%% Fig.1  %%%%% model for March 7
%\begin{figure}[t]
%\epsfxsize=20cm % fix the x-dimension and scales y-dim. to x-dim.
%\hspace{-3.cm}
%\plotone{sn97ef7mar.ps}
\plotone{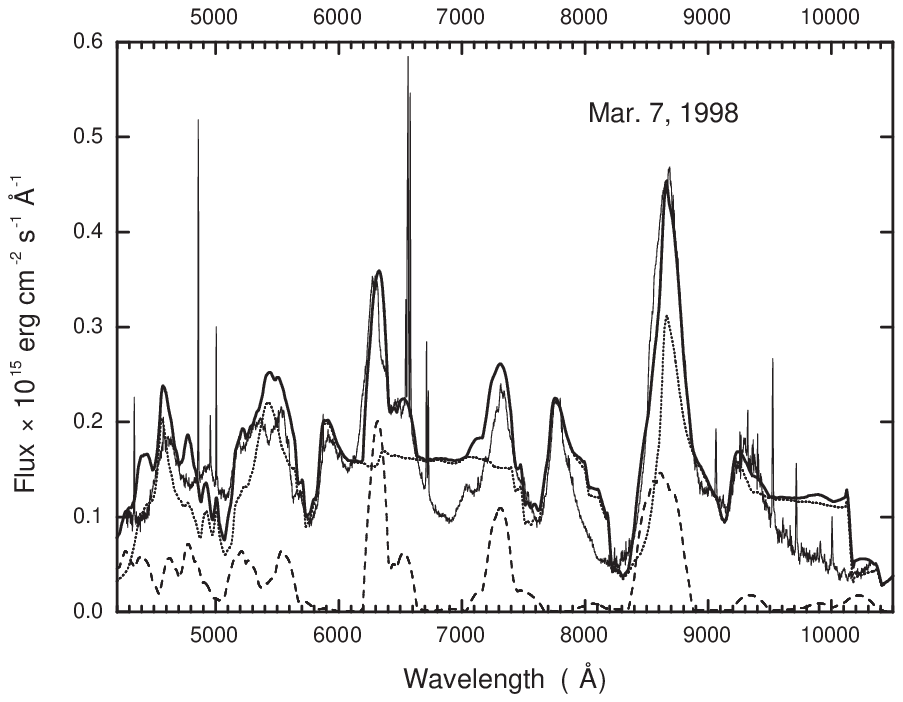}
\figcaption[mazzali_f1.eps]{Spectrum of SN~1997ef, obtained on 1998 March 7 (solid line),
and comparison with synthetic spectra: the photospheric component (dotted
line), the nebular component (dashed line), and their sum (thick solid line).}
%\end{figure}
%\clearpage

%%%%%% Fig.2   %%%%% model for March 26
%\begin{figure}[t]
%\epsfxsize=20cm % fix the x-dimension and scales y-dim. to x-dim.
%\hspace{-2.cm}
%\plotone{sn97ef26mar.ps}
\plotone{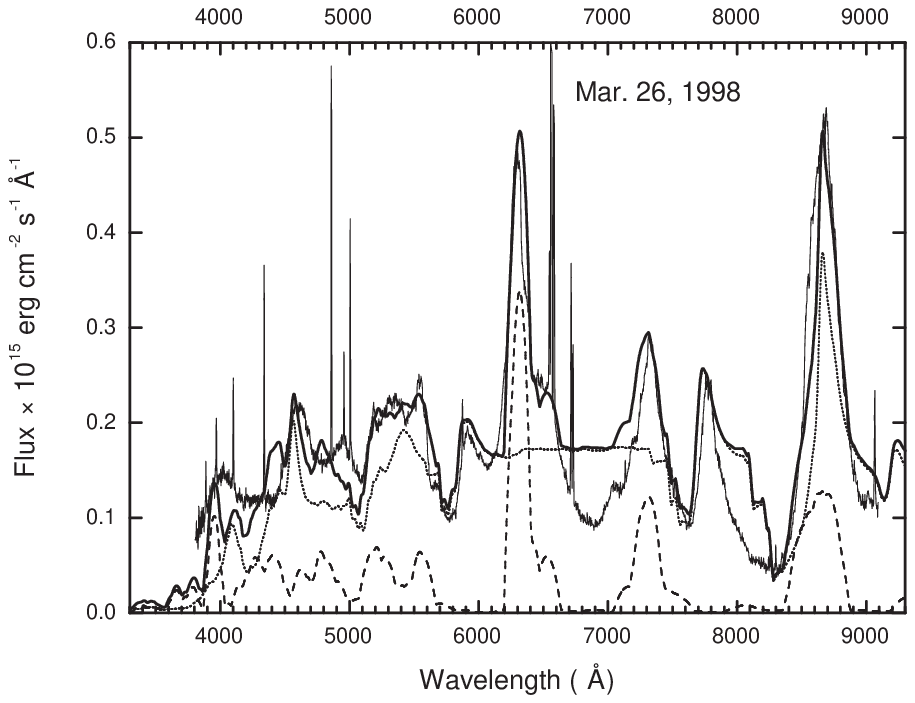}
\figcaption[mazzali_f2.eps]{Spectrum of SN~1997ef, obtained on 1998 March 26 (solid 
line), and comparison with synthetic spectra: the photospheric component (dotted
line), the nebular component (dashed line), and their sum (thick solid line).}
%\end{figure}
%\clearpage

%%%%%% Fig.3   %%%%% Light curve of SN~1997ef
%\begin{figure}[t]
%\epsfxsize=17cm % fix the x-dimension and scales y-dim. to x-dim.
%\hspace{.3cm}
%\plotone{/home/mazzali/gamma/sn97efLClate.ps}
\plotone{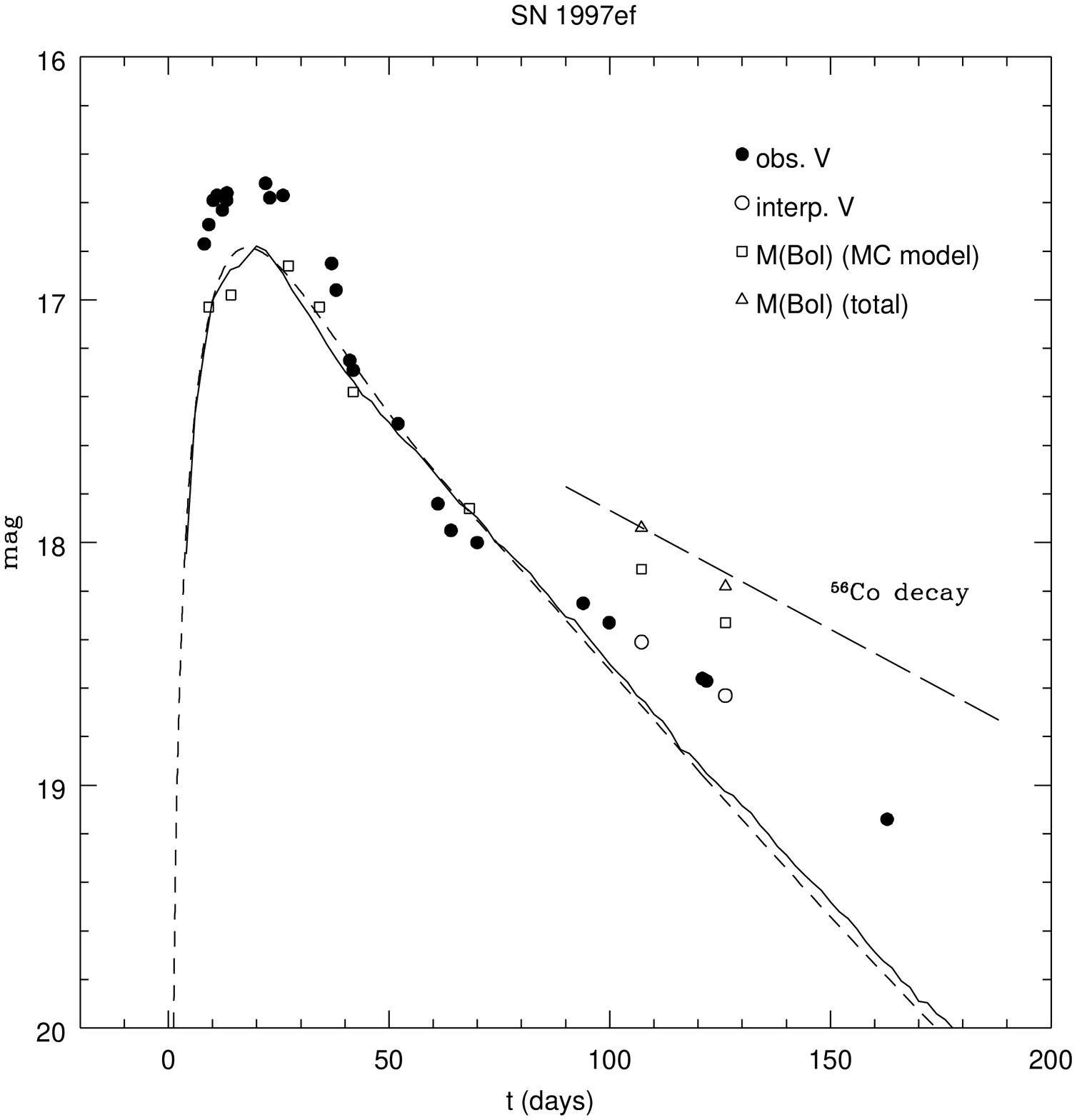}
\figcaption[mazzali_f3.eps]{Light curve of SN~1997ef. The smooth lines are synthetic
bolometric light curves: the dashed curve was computed with a radiation
hydrodynamics code,
while the solid one was computed with a Monte Carlo code (Mazzali \etal 2000).
Both synthetic light curves were computed using the density distributions
derived from reproducing the spectra. }
%\end{figure}
%\clearpage

%%%%%% Fig.4   %%%%% Light curves of SN~1997ef and SN~1997dq
%\begin{figure}[t]
%\epsfxsize=17cm % fix the x-dimension and scales y-dim. to x-dim.
%\hspace{.3cm}
%\plotone{sn97ef97dqlc.ps}
\plotone{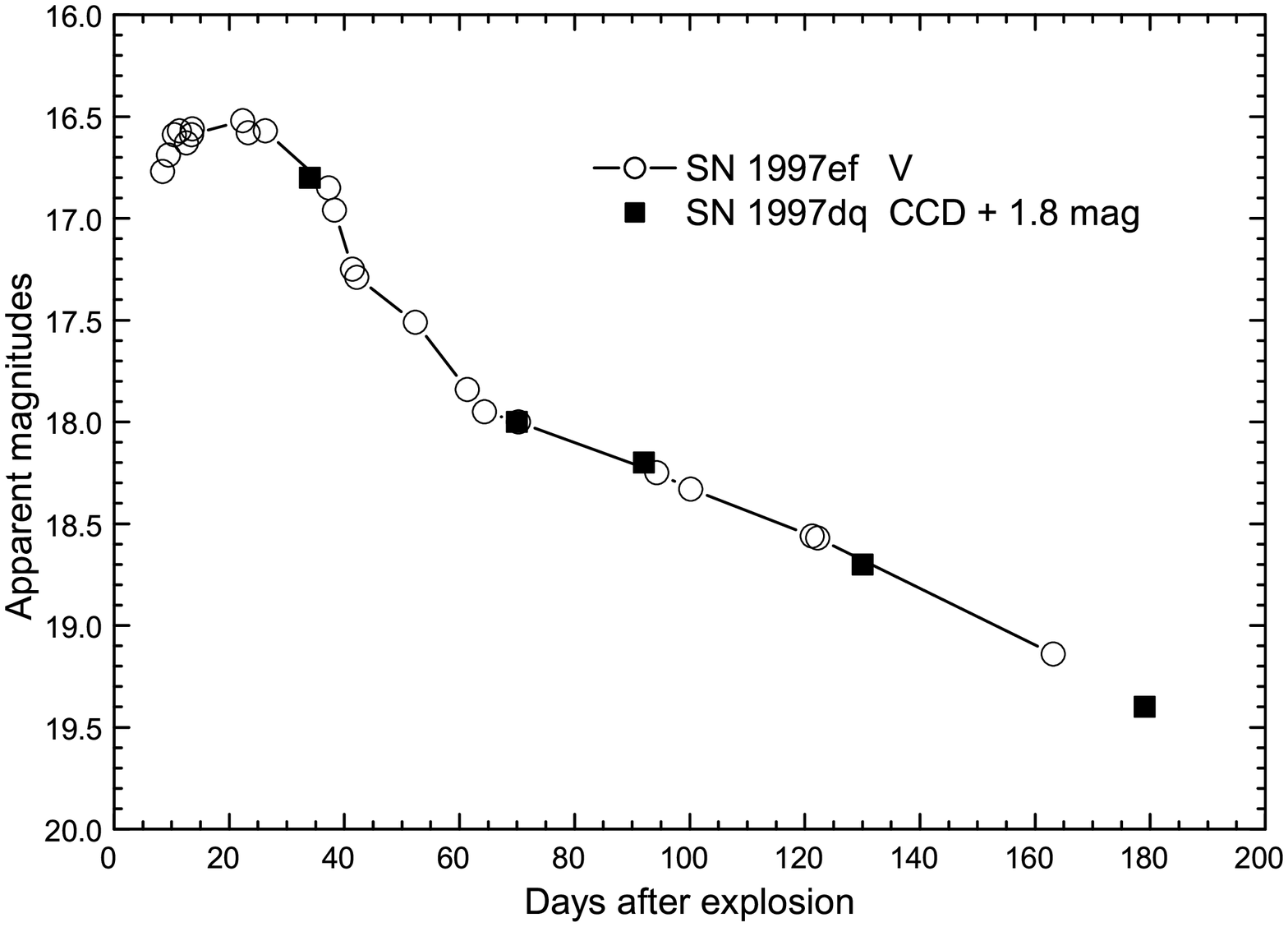}
%for centering: act on hspace argument
\figcaption[mazzali_f4.eps]{The $V$-band light curve of SN~1997ef compared to the CCD 
light curve of SN~1997dq. The latter has been shifted down by 1.8 mag and 
shifted along the time axis to match the reference SN~1997ef light curve.}
%\end{figure}
%\clearpage

%%%%%% Fig.5   %%%%% Spectra of SN~1997ef and SN~1997dq
%\begin{figure}[t]
%\epsfxsize=17cm % fix the x-dimension and scales y-dim. to x-dim.
%\hspace{.3cm}
%\plotone{/home/mazzali/newmc/sne97ef97dqcomp.eps}
\plotone{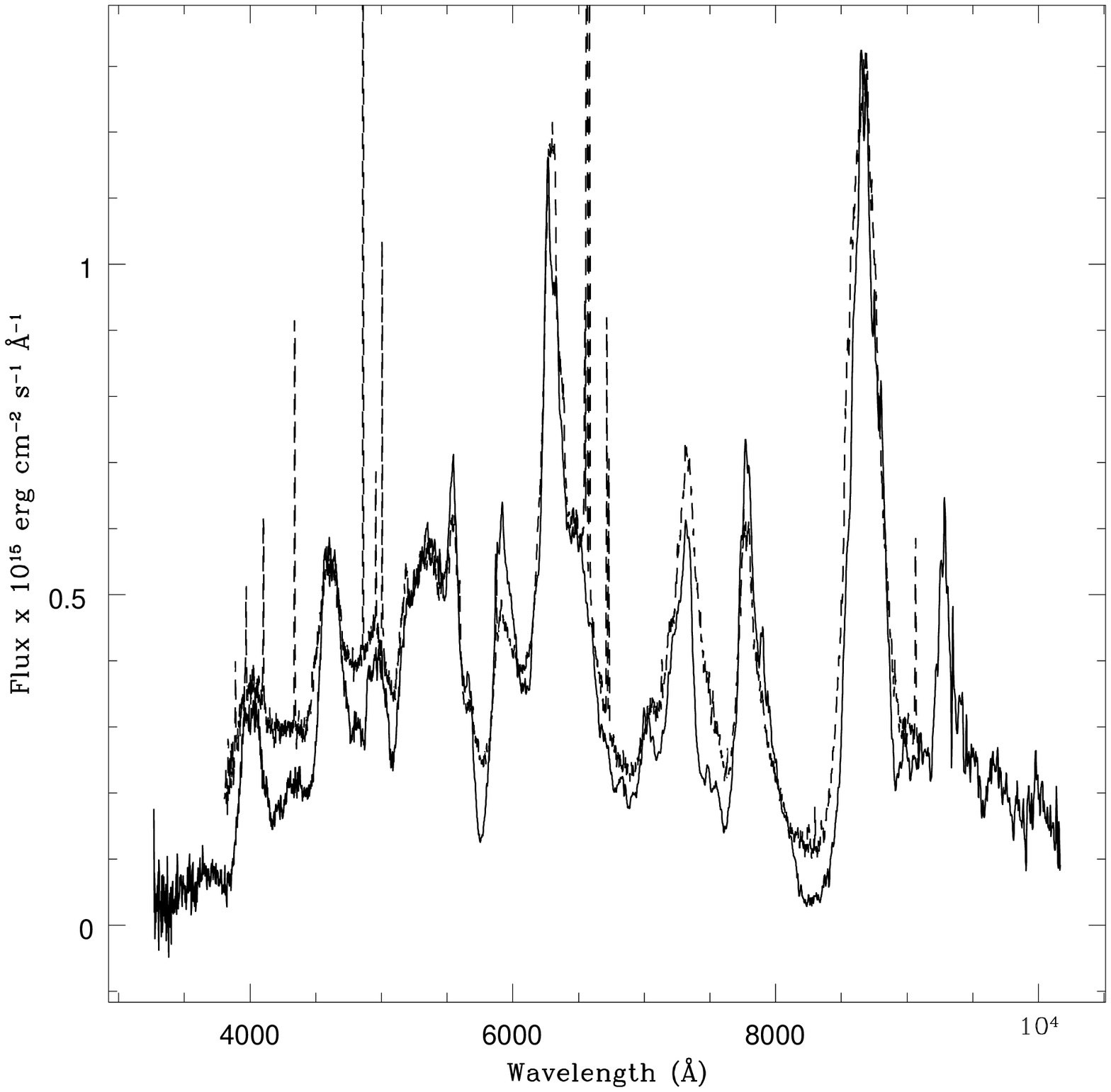}
\figcaption[mazzali_f5.eps]{The spectrum of SN~1997ef taken on 1998 Jan. 26 (dashed 
line), 126 days after the estimated time of outburst, compared to the spectrum 
of SN~1997dq obtained on 1998 Jan. 28. The epoch of this spectrum is $\sim 126$
days based on the comparison of the light curves of the two SNe. The flux of
the spectrum of SN~1997ef has been multiplied by a factor of 2.5 for clarity.}
%\end{figure}
%\clearpage

%%%%%% Fig.6   %%%%% Nebular model for SN~1997dq
%\begin{figure}[t]
%\epsfxsize=17cm % fix the x-dimension and scales y-dim. to x-dim.
%\hspace{.3cm}
%\plotone{/home/mazzali/neb/neb97dqd262mu31.83.eps}
\plotone{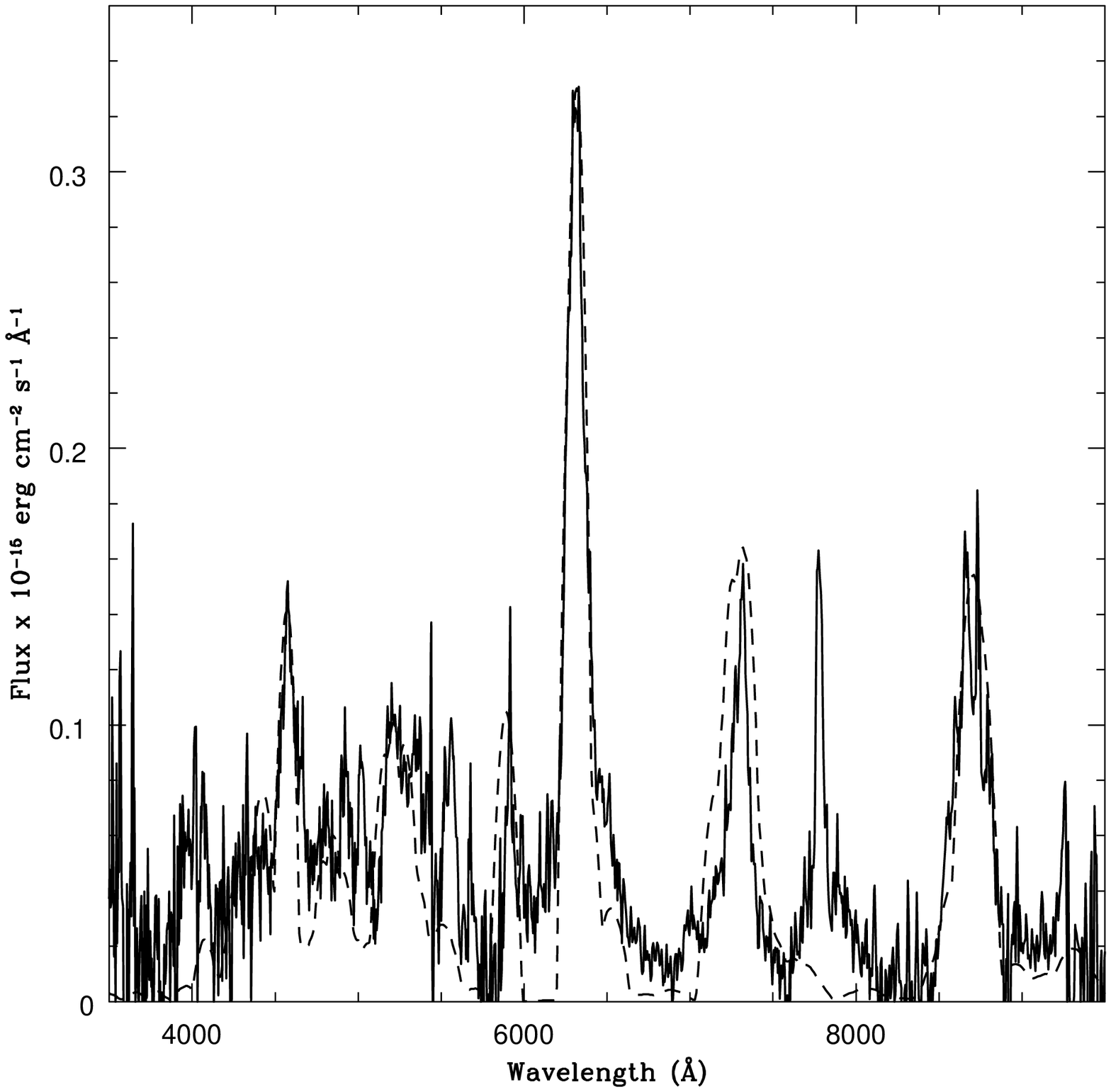}
\figcaption[mazzali_f6.eps]{The spectrum of SN~1997dq taken on 1998 June 18, 262 days 
after the estimated time of outburst, compared to a synthetic nebular spectrum
computed with our NLTE code (dashed line).}
%\end{figure}
%\clearpage

\end{document}